\documentclass[a4paper]{panle}
\usepackage{url}
\usepackage{cite}
\usepackage{wrapfig}
\usepackage{graphicx}
\usepackage{amssymb}
\usepackage{amsfonts}
\usepackage{amsmath}
\usepackage{longtable}
\usepackage{rotating}
\usepackage{lscape}
\usepackage{epsfig}
\usepackage{multirow}
\usepackage{bm}

\usepackage{hyperref}

\originalTeX
\begin{document}

\title{Directed Flow of Protons and Deuterons in Xe+Cs(I) Collisions: Preliminary BM@N Data and THESEUS Modeling}
\maketitle
\authors{M.\,E.~Kozhevnikova\,$^{a,}$\footnote{E-mail: kozhevnikova@jinr.ru},
	M.\,V.~Mamaev\,$^{a,b}$,
	I.\,A.~Zhavoronkova\,$^{a,b}$,
	A.\,V.~Taranenko\,$^{a,b}$,
	Yu.\,B.~Ivanov\,$^{a,c}$}
\setcounter{footnote}{0}
\from{$^{a}$\, Joint Institute for Nuclear Research, Dubna, Moscow Region, Russia, 141980}

\from{$^{b}$\,National Research Nuclear University "MEPhI", Moscow, Russia, 115409}

\from{$^{c}$\,National Research Centre "Kurchatov Institute", Moscow, Russia, 123182}



\begin{abstract}

Preliminary BM@N results on the directed flow ($v_1$) of protons and deuterons in Xe+Cs(I) collisions at 3.8$A$ GeV are presented for the 10–40\% centrality interval. 
The measured rapidity dependence of $v_1$ is compared with calculations from the THESEUS event generator, where deuterons are produced thermodynamically on an equal basis with hadrons using a late freeze-out scenario. 
While THESEUS well describes the proton $v_{1}$ data, it shows a slight but systematic overestimation of the deuteron flow at low and intermediate rapidities. 
This comparison tests both the collective dynamics of baryon-rich matter and the thermodynamic mechanism of light-nucleus formation at Nuclotron energies.

\end{abstract}
\vspace*{6pt}


\label{sec:intro}
\section*{INTRODUCTION}

Directed flow is a sensitive probe of the early-stage dynamics of nuclear matter produced in relativistic heavy-ion collisions. 
At Nuclotron energies, it reflects the collective dynamics of baryon-rich matter and thus provides information on the pressure generated during the collision, which, in turn, offers key insights into the Equation of State (EoS).
The comparison of proton and deuteron flow can test both the collision dynamics and the mechanism of light-nucleus formation.
To quantify these collective phenomena, the azimuthal distribution of produced particles relative to the reaction-plane angle $\Psi_{RP}$
is conventionally expressed as a Fourier series,
\begin{equation*}
	\frac{dN}{d\varphi}	= \frac{N}{2\pi} \left[1+2\sum_{n=1}^{\infty}	v_n\cos\left(n(\varphi-\Psi_{\rm RP})\right)\right],
	\label{eq:fourier}
\end{equation*}
where $\varphi$ is the particle azimuthal angle and $v_n$ are the anisotropic flow coefficients~\cite{Voloshin:2008dg}. 
The first one, $v_1= \langle \cos(\varphi-\Psi_{\rm RP}) \rangle$,
is referred to as directed flow. 

The BM@N (Baryonic Matter at Nuclotron) experiment at the NICA  probes the region of high baryon densities in relativistic nucleus-nucleus collisions.
In this paper, preliminary BM@N data on the directed flow of protons and deuterons from Xe+Cs(I) collisions at a beam energy of 3.8$A$~GeV are presented for the 10–40\% centrality interval. These experimental results are analyzed within the hybrid event generator THESEUS (Three-fluid Hydrodynamics-based Event Simulator Extended by UrQMD final State interactions)~\cite{Kozhevnikova:2020bdb}, where UrQMD stands for Ultra-relativistic Quantum Molecular Dynamics. 


\label{sec:model}
\section{MODEL}

The THESEUS event generator couples the Three-Fluid Fynamics model (3FD)~\cite{Ivanov:2005yw} with a Monte-Carlo particle-sampling procedure and the UrQMD transport afterburner for hadrons. The 3FD stage simulates baryon stopping and nonequilibrium expansion using three interacting fluids. In this analysis, a smooth crossover EoS is employed for the hydrodynamic evolution. 
At particlization, fluid elements are converted into an ensemble of particles via sampling from local thermal distributions. Hadrons subsequently undergo final-state interactions in UrQMD. In this updated version of THESEUS, stable light nuclei are included directly in the thermodynamic particle table~\cite{Kozhevnikova:2020bdb}. Deuterons are thus sampled at freeze-out on an equal basis with hadrons, and their abundance is determined by local thermodynamic conditions rather than by a nucleon-coalescence procedure. 

Currently, light nuclei do not participate in the UrQMD cascade. To approximate their hadronic-stage interactions, a later freeze-out criterion with a universal energy density of $\varepsilon_{\rm frz} = 0.2\,{\rm GeV/fm^3}$ is applied~\cite{Kozhevnikova:2023mnw}, and is not specifically adjusted for the deuteron flow.

\section{PRELIMINARY BM@N DATA}

BM@N is a fixed-target magnetic spectrometer at the Nuclotron/NICA complex (JINR). Its tracking system consists of GEM and Silicon detectors that measure charged-particle momenta, the time-of-flight (ToF) system that provides hadron identification, and the Forward Hadron Calorimeter (FHCal) that reconstructs the spectator symmetry plane.

Preliminary data were obtained for Xe+Cs(I) collisions at kinetic beam energy of 3.8$A$~GeV. Protons and deuterons were identified via mass-squared ($m^2$) from the ToF. The directed flow ($v_1$) was measured relative to the first-order spectator plane (for further information see Ref.~\cite{Mamaev:2024wiy, Zhavoronkova:2026mlf}). Data correction and the three-sub-event method were applied to account for the event-plane resolution. 
The analysis~\cite{Segal:2023njv} was performed for the 10--40\% centrality interval, ensuring optimal resolution and a strong $v_1$ signal.



\section{RESULTS AND DISCUSSION}

\begin{figure}[htb]
	\begin{center}
		\includegraphics[width=135mm]{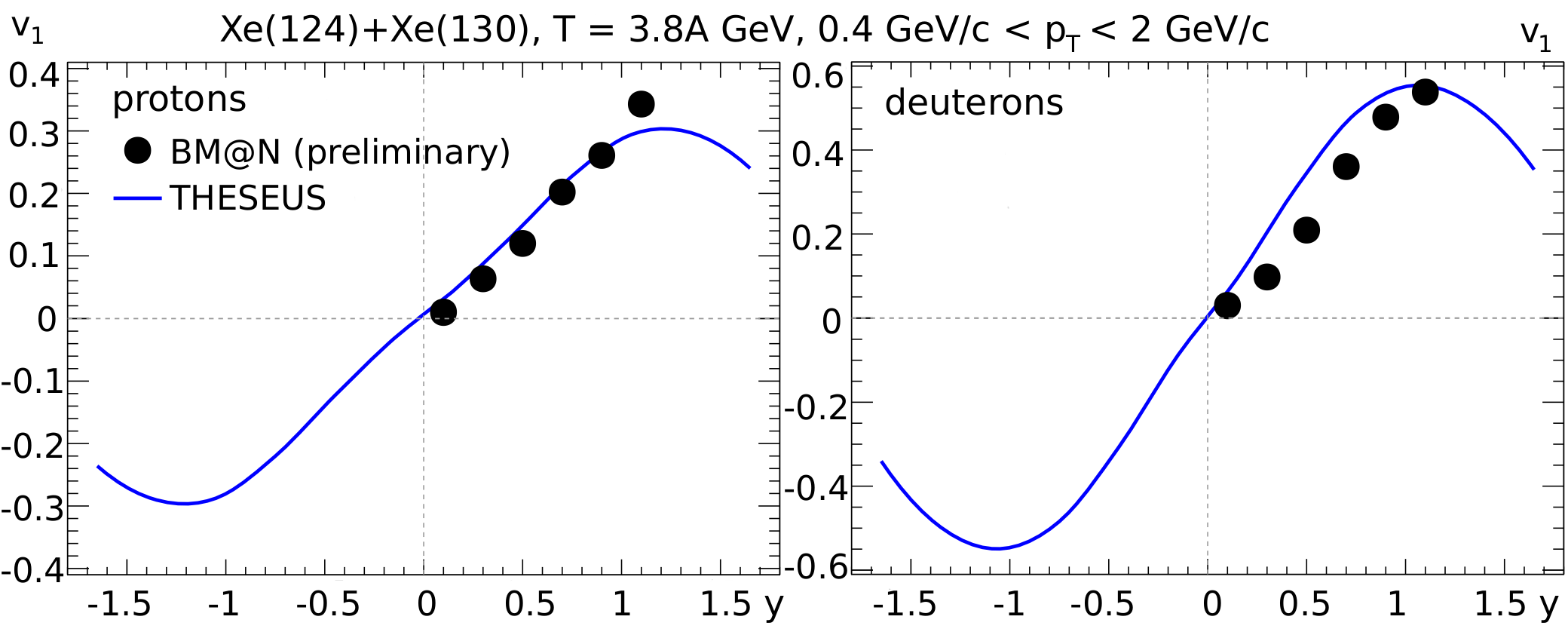}
		\vspace{-3mm}
		\caption{THESEUS calculation (blue line) of directed flow $v_1$ for protons (left panel) and deuterons (right panel) in comparison with preliminary BM@N data (black dots). In 3FD/THESEUS the crossover EoS is used.}
		\label{fig:v1}
	\end{center}
	\vspace{-5mm}
\end{figure}

Fig.~\ref{fig:v1} presents the directed flow $v_1$ of protons and deuterons as a function of the rapidity $y$ in collisions of Xe(124)+Xe(130) which are used to approximate the Xe+Cs(I) system due to their similar mass and charge numbers and with the kinetic energy $T = 3.8A$~GeV.  Experimental cuts on transverse momentum $0.4\,{\rm GeV/c} <p_T < 2.0\,{\rm GeV/c}$ are applied.

As in 3FD/THESEUS only one value of impact parameter is available for one run of calculations, 
we calculate the average $v_1$ in the proper centrality interval of impact parameters similarly to Eq.~(15) from Ref.~\cite{Kolb_2001}. 
Since the results are preliminary, this comparison is intended primarily as a test of the general trends of the proton and deuteron flow.
It is seen that the EoS with smooth crossover in 3FD/THESEUS gives a good description of the data of protons by THESEUS in the rapidity region of $|y|\lesssim 1$.
The situation with deuterons is somewhat worse: the model slightly overesimates the BM@N data and the calculated midrapidity slope of the deuteron $v_1$ is slightly steeper than the experimental one, however, this is in agreement with previous THESEUS studies at $\sqrt{s_{NN}} = 3$~GeV \cite{Kozhevnikova:2023mnw}.

\section*{CONCLUSIONS}
We compared preliminary BM@N data on proton and deuteron directed flow in Xe+Cs(I) collisions at 3.8\(A\)~GeV with THESEUS calculations using a crossover EoS. The model reproduces the main trends of the measured flow. While the calculated deuteron flow slightly overestimates the experimental data, the proton flow is well described. These results support the thermodynamic approach to light-nucleus production in THESEUS and highlight the potential of joint proton-deuteron flow analyses at BM@N energies.

\section*{FUNDING}
The work was funded by the Ministry of Science and Higher Education of the Russian Federation, Project "Studying physical phenomena in the micro- and macro-world to develop future technologies" FSWU-2026-0010

\section*{ACKNOWLEDGEMENTS}
This work was carried out using computing resources of the federal collective usage center "Complex for simulation and data processing for megascience facilities" at NRC "Kurchatov Institute" and computing resources of the supercomputer "Govorun" at JINR.

\section*{CONFLICT OF INTEREST}

The authors declare that they have no conflicts of interest.


\bibliographystyle{pepan}
\bibliography{pepan_biblio}

\end{document}